\documentclass[aps,prc,reprint,superscriptaddress,amsmath]{revtex4-1}

\usepackage{amssymb}
\usepackage{hyperref}
\hypersetup{
  bookmarksopen=true,
  pdftitle={Quantifying properties of hot and dense QCD matter through systematic model-to-data comparison}
}
\usepackage{graphicx}
\graphicspath{{fig/}}
\usepackage{multirow}
\usepackage{color}
\usepackage{appendix}

\newcommand{\colfig}[3][t]{
  \begin{figure}[#1]
    \includegraphics{#2}
    \caption{\label{fig:#2}#3}
  \end{figure}
}

\newcommand{\widefig}[3][t]{
  \begin{figure*}[#1]
    \includegraphics{#2}
    \caption{\label{fig:#2}#3}
  \end{figure*}
}

\newcommand{\avg}[1]{\langle #1 \rangle}
\newcommand{\nch}{N_\text{ch}}
\newcommand{\vnk}[2]{v_#1\{#2\}}
\newcommand{\tran}{^\intercal}
\newcommand{\trento}{T\raisebox{-.5ex}{R}ENTo}
\newcommand{\order}[1]{$\mathcal O(10^{#1})$}
\newcommand{\paddedhline}{\noalign{\smallskip}\hline\noalign{\smallskip}}

\begin{document}

\title{Quantifying properties of hot and dense QCD matter \\ through systematic model-to-data comparison}

\author{Jonah E.\ Bernhard}
\affiliation{Department of Physics, Duke University, Durham, NC 27708}

\author{Peter W.\ Marcy}
\affiliation{Statistical Sciences Group, Los Alamos National Laboratory, Los Alamos, NM 87545}

\author{Christopher E.\ Coleman-Smith}
\affiliation{Department of Physics, Duke University, Durham, NC 27708}

\author{Snehalata Huzurbazar}
\affiliation{Department of Statistics, University of Wyoming, Laramie, WY 82071}

\author{Robert L.\ Wolpert}
\affiliation{Department of Statistical Science, Duke University, Durham, NC 27708}

\author{Steffen A.\ Bass}
\affiliation{Department of Physics, Duke University, Durham, NC 27708}

\date{\today}

\begin{abstract}
  We systematically compare an event-by-event heavy-ion collision model to data from the Large Hadron Collider.
  Using a general Bayesian method, we probe multiple model parameters including fundamental quark-gluon plasma properties such as the specific shear viscosity $\eta/s$, calibrate the model to optimally reproduce experimental data, and extract quantitative constraints for all parameters simultaneously.
  The method is universal and easily extensible to other data and collision models.
\end{abstract}

\maketitle

\section{Introduction}

Relativistic heavy-ion collisions produce a hot, dense phase of strongly-interacting matter commonly known as the quark-gluon plasma (QGP), which rapidly expands and freezes into hadrons \cite{Arsene:2004fa,Adcox:2004mh,Back:2004je,Adams:2005dq,Gyulassy:2004zy,Muller:2006ee,Muller:2012zq}.
Since the QGP is not directly observable---only final-state hadrons are detected---present research seeks to quantify the fundamental properties of the QGP, such as its transport coefficients and the nature of the initial state, through comparisons of experimental measurements to computational model calculations.

Computational models must take a set of input parameters including the physical properties of interest, simulate the full time-evolution of heavy-ion collisions, and produce outputs analogous to experimental measurements.
The true values of the physical properties are extracted by calibrating the input parameters so that the model output optimally reproduces experimental data.
This generic recipe is called ``model-to-data comparison''.

Notably, the QGP shear viscosity to entropy density ratio $\eta/s$ has been constrained by comparing anisotropic flow coefficients $v_n$ between model and experiment.
Explicit calculation of $\eta/s$ directly from QCD is not yet feasible, and while there is a conjectured lower bound $\eta/s~\geq~1/4\pi~\simeq~0.08$ from AdS/CFT holography \cite{Kovtun:2004de}, model-to-data comparison is the most attractive option for determining the optimal input parameter value and corresponding uncertainty.
To this end, previous studies used viscous relativistic fluid dynamics and hybrid transport models to compute $v_n$ at several values of $\eta/s$, then chose the value which most closely matched experimental $v_n$.
A variety of complementary calculations have constrained $\eta/s$ to an approximate range of 0.08--0.20 \cite{Luzum:2008cw,Song:2010mg,Schenke:2010rr,Luzum:2012wu}.

However, $\eta/s$ is not the only model input parameter:
many other parameters remain poorly determined, e.g.~the hydrodynamic thermalization time $\tau_0$ and initial conditions; and models often have non-physical nuisance parameters that nonetheless should be tuned to optimal values.
The flow coefficients $v_n$ are only a small subset of all QGP observables:
models must also describe basic quantities such as the charged-particle multiplicity and transverse-momentum distributions.

Recent work \cite{Soltz:2012rk} moved toward a more global analysis of multiple model parameters and observables, but encountered practical limitations attempting to simultaneously tune these free parameters.
In general, input parameters correlate among each other and contribute to multiple observables, so they cannot be constrained independently.

Algorithms such as Markov chain Monte Carlo (MCMC) can rigorously explore this type of complex high-dimensional parameter space, but require a very large number of model evaluations---often thousands or millions, depending on the problem at hand.
Heavy-ion collision models may run for several hours, so a direct MCMC approach is intractable.
The situation is exacerbated when studying event-by-event fluctuations as opposed to average quantities:
while event-averaged models save computation time by using a smooth initial condition and single hydrodynamic calculation, event-by-event models have realistic, fluctuated initial conditions, each of which requires its own hydrodynamic treatment.
Many thousands of complete events are necessary \emph{at each point in parameter space} to capture event-by-event fluctuations.

These limitations may be overcome through a modern Bayesian method for analyzing computationally expensive models \cite{OHagan:2006ba,Higdon:2008cmc,Higdon:2014tva}.
A set of salient model parameters is chosen for calibration---the set should include any fundamental physical properties of interest---and the model is evaluated at a relatively small \order 2 number of points.
Those points are then interpolated with a Gaussian process emulator \cite{Rasmussen:2006gp} to provide a continuous picture of the parameter space.
The emulator acts as a fast surrogate to the full model:
it predicts model output at arbitrary points in parameter space with negligible computational cost.
This effectively removes most practical barriers and enables parameter calibration through standard techniques such as MCMC.

Emulators have been successfully used to study a wide range of physical systems, including galaxy formation \cite{Gomez:2012ak} and heavy-ion collisions \cite{Novak:2013bqa,Pratt:2014cza,Pratt:2015zsa}.
Reference \cite{Novak:2013bqa} calibrated a hydrodynamic model to identified particle spectra from the Relativistic Heavy Ion Collider (RHIC) and extracted constraints on $\eta/s$ and several initial state parameters.
However, this study used an event-averaged initial condition model, limiting its ability to investigate event-by-event fluctuations.

In this work, we apply Bayesian methodology to a full event-by-event heavy-ion collision model.
We calibrate to multiplicity and flow data from the Large Hadron Collider (LHC) and constrain the shear viscosity $\eta/s$ along with other hydrodynamic and initial condition parameters.
The analysis framework handles arbitrary numbers of inputs and outputs, systematically calculates quantitative constraints on all inputs simultaneously, and quickly evaluates the efficacy of physical models.

\section{Model}

State-of-the-art heavy-ion collision models simulate QGP spacetime evolution in several stages \cite{Bass:2000ib,Teaney:2001av,Hirano:2005xf,Nonaka:2006yn,Petersen:2008dd,Song:2010mg,Schenke:2010rr,Shen:2014vra}:
\begin{enumerate}
  \item an initial condition model describes the initial state and non-equilibrium dynamics until QGP formation,
  \item viscous relativistic hydrodynamics calculates the dynamical expansion of the hot and dense QGP medium including the phase transition to a hadron gas,
  \item then a particlization model converts the system into a microscopic ensemble of hadrons,
  \item and finally a Boltzmann transport model calculates hadronic rescattering and decays.
\end{enumerate}
In this work, we opt for a mature, well-tested set of event-by-event models \cite{Shen:2014vra} with an established track record of describing diverse RHIC and LHC data \cite{Song:2010mg,Shen:2011zc,Song:2011hk}.
This choice will permit direct comparison between existing results and the outcome of the following systematic model-to-data comparison.
We emphasize, however, that the methodology in this paper can easily be applied to any set of models and corresponding data.

\subsection{Initial conditions}

Initial condition models provide the outcome of the collision's pre-equilibrium evolution at the hydrodynamic thermalization time, approximately 0.5 fm/$c$.
Some models explicitly calculate pre-equilibrium dynamics \cite{Schenke:2012wb} starting from the initial state of the collision; others skip this time frame and generate initial conditions directly at the thermalization time \cite{Drescher:2006pi,Miller:2007ri,Moreland:2014oya}.

We select two of the most widely used models in the latter category:
the Monte Carlo Glauber \cite{Miller:2007ri} and Monte Carlo KLN \cite{Drescher:2006pi} models.
Although more sophisticated models were recently introduced \cite{Schenke:2012wb,Moreland:2014oya}, both Glauber and KLN provide reasonable event-by-event initial conditions with well-understood behavior and a broad basis of published results.

\subsection{Hydrodynamics}

The initial condition furnishes the hydrodynamic stress-energy tensor $T^{\mu\nu}$ at the thermalization time $\tau_0$.
Viscous hydrodynamics then solves the conservation equations
\begin{equation}
  \partial_\mu T^{\mu\nu} = 0
\end{equation}
where
\begin{equation}
  T^{\mu\nu} = (\epsilon + P) u^\mu u^\nu - P g^{\mu\nu} + \pi^{\mu\nu};
\end{equation}
$\epsilon$, $P$, and $u^\mu$ are the energy density, pressure, and flow velocity of the fluid; $g^{\mu\nu}$ is the metric tensor; and $\pi^{\mu\nu}$ is the shear stress tensor.
An equation of state
\begin{equation}
  P = P(\epsilon)
\end{equation}
closes the system of hydrodynamic equations and is usually provided by a parametrization of lattice QCD calculations.

We employ an improved version of \small{VISH2+1} \cite{Song:2007ux}, a stable, extensively tested implementation of boost-invariant viscous hydrodynamics that was recently updated to handle fluctuating event-by-event initial conditions \cite{Shen:2014vra}.
\small{VISH2+1} uses the prevalent s95 partial chemical equilibrium equation of state \cite{Huovinen:2009yb}.

\subsection{Particlization}

As the hydrodynamic medium expands and cools, it undergoes a transition from a deconfined QGP to a hot and dense hadronic system.
At this point it's advantageous to switch to a microscopic transport model, for such models naturally account for the system's increasing viscosity, non-equilibrium break-up, and eventual freeze-out.
Thus, a particlization model converts the fluid into a microscopic ensemble of hadrons once the fluid cools to a pre-specified switching temperature, typically just below the QCD transition temperature $T_c \sim 165$ MeV.
The model generates particles by sampling the Cooper-Frye formula \cite{Cooper:1974mv}
\begin{equation}
  E \frac{dN_i}{d^3p} = \int_\sigma f_i(x,p) \, p^\mu \, d^3\sigma_\mu,
\end{equation}
where $f_i$ is the distribution function for particle species $i$, $p^\mu$ is the four-momentum, and the integral is taken over the isothermal spacetime hypersurface $\sigma$ defined by the switching temperature.

We use a recent hypersurface sampler designed to couple with \small{VISH2+1} \cite{Qiu:2013wca,Shen:2014vra}.

\subsection{Hadronic phase}

After particlization, the medium continues to expand as an interacting hadron gas (e.g.\ scatterings and decays).
A hadronic ``afterburner'' calculates these interactions through the Boltzmann equation
\begin{equation}
  \frac{df_i(x,p)}{dt} = \mathcal C_i(x,p),
\end{equation}
where $f_i$ is the distribution function and $\mathcal C_i$ is the collision kernel which contains all possible hadronic interactions for particle species $i$.
Particles emerging from the afterburner are analogous to particles streaming into an experimental detector.

We adopt Ultra-relativistic Quantum Molecular Dynamics (UrQMD) \cite{Bass:1998ca,Bleicher:1999xi} as an afterburner.

\subsection{Postprocessing}

The full event-by-event model is executed \order 4 times for each set of input parameters, yielding \order 7 events in total.
Events are binned into centrality intervals and the raw event data are postprocessed into physical observables for direct comparison with experiment.
In this analysis we calculate the centrality dependence of several standard observables:
the average charged-particle multiplicity $\avg\nch$ and multi-particle flow cumulants $\vnk n {2k}$.
Note that the method trivially extends to arbitrary numbers and types of observables---all that's required is a model calculation and a corresponding experimental measurement.

Flow cumulants $\vnk n {2k}$ are defined as the $2k$-particle correlation function of the $n$th-order azimuthal anisotropy.
For example, the two-particle cumulant is
\begin{equation}
  \vnk n 2^2 \equiv \bigl\langle e^{in(\phi_i - \phi_j)} \bigr\rangle,
\end{equation}
where $\phi_i$ is the azimuthal angle of the transverse momentum of particle $i$ and the average is over all distinct pairs of particles $i,j$.
The two-particle cumulant is also approximately equal to the root-mean-square of the full $v_n$ distribution \cite{Borghini:2000sa}:
\begin{equation}
  \vnk n 2 \simeq \sqrt{\avg{v_n^2}}.
\end{equation}
We compute two-particle cumulants for elliptic and triangular flow $\vnk 2 2$, $\vnk 3 2$ using the direct $Q$-cumulant method \cite{Bilandzic:2010jr}.
Higher-order cumulants are currently out of reach due to insufficient quantities of events.

Postprocessed observables are compared to corresponding experimental results recently measured by the ALICE experiment at the LHC for Pb-Pb collisions at $\sqrt{s_\text{NN}} = 2.76$ TeV \cite{Abelev:2014mda}.
All observables are subjected to the same kinematic cuts as the ALICE detector, namely charged particles with $|\eta| < 1$ and $0.2 < p_T < 3.0$ GeV.

\section{Emulator}

This section constructs a Gaussian process (GP) emulator to act as a surrogate for the full event-by-event model.
The strategy is to evaluate the model on a carefully chosen set of input parameter points, then use a GP to interpolate the parameter space.
Unlike alternative interpolation schemes such as splines or polynomial interpolation, a GP emulator provides a \emph{probability distribution} at each point in parameter space, hence, it not only predicts the output of the model at arbitrary points in parameter space, but also quantifies the uncertainty of its predictions.
Further, GPs are non-parametric interpolators, i.e.\ they do not require an assumed functional form for the underlying model.
These features are essential for emulation of computer codes.

\subsection{Gaussian processes}

\newcommand{\x}{\mathbf x}
\newcommand{\y}{\mathbf y}
\newcommand{\zero}{\mathbf 0}
\newcommand{\muvec}{\boldsymbol\mu}
\newcommand{\N}{\mathcal N}

This subsection summarizes the theory of Gaussian process emulators as detailed in Chap.~2 of \cite{Rasmussen:2006gp}.

A Gaussian process (GP) is defined as a collection of random variables, any finite number of which have a joint Gaussian distribution.
A GP may be thought of as a stochastic function $f(\x)$ which maps $n$-dimensional input vectors $\x$ to normally distributed outputs $y$.
It is fully specified by a mean function $\mu(\x)$ which gives the mean of $f$ at input point $\x$ and a covariance function $\sigma(\x, \x')$ which provides the covariance of $f$ between a pair of points $\x$, $\x'$.

As a concrete example, let $\x_1$ be an input point and $y_1~=~f(\x_1)$ be the output of the GP at $\x_1$; then $y_1$ has a normal distribution with mean $\mu(\x_1)$ and variance $\sigma(\x_1, \x_1)$:
\begin{equation}
  y_1 \sim \N(\mu(\x_1), \sigma(\x_1, \x_1)).
\end{equation}
Now if $\x_2$ is another input and $y_2 = f(\x_2)$ is the corresponding output, $y_1$ and $y_2$ have a bivariate normal distribution
\begin{equation}
  \begin{pmatrix}
    y_1 \\ y_2
  \end{pmatrix}
  \sim \N\biggl[
    \begin{pmatrix}
      \mu(\x_1) \\ \mu(\x_2)
    \end{pmatrix},
    \begin{pmatrix}
      \sigma(\x_1, \x_1) & \sigma(\x_1, \x_2) \\
      \sigma(\x_2, \x_1) & \sigma(\x_2, \x_2) \\
    \end{pmatrix}
  \biggr].
\end{equation}
In general, the set of $m$ random output variables $\y~=~\{y_1, \ldots, y_m\}~=~f(X)$ corresponding to input points $X~=~\{\x_1, \ldots, \x_m\}$ have a multivariate normal distribution
\begin{equation}
  \y \sim \N(\muvec, \Sigma)
\end{equation}
where
\begin{equation}
  \muvec = \mu(X) = \{\mu(\x_1), \mu(\x_2), \ldots, \mu(\x_m)\} \\
\end{equation}
is the $m$-dimensional mean vector from applying the mean function to each input, and
\begin{equation}
  \Sigma = \sigma(X, X) =
  \begin{pmatrix}
    \sigma(\x_1, \x_1) & \cdots & \sigma(\x_1, \x_m) \\
    \vdots & \ddots & \vdots \\
    \sigma(\x_m, \x_1) & \cdots & \sigma(\x_m, \x_m) \\
  \end{pmatrix}
\end{equation}
is the $m \times m$ covariance matrix from applying the covariance function to each pair of inputs.

In practice, the mean function is often set to zero, since the mean of a distribution can always be subtracted off.
The covariance function must be carefully chosen, for it controls the degree of similarity between pairs of points.
A standard choice is the squared-exponential function
\begin{equation}
  \sigma(\x, \x') = \exp\biggl( -\frac{|\x - \x'|^2}{2\ell^2} \biggr),
\end{equation}
where $\ell$ is a characteristic length scale.
Notice that nearby points are strongly correlated ($\sigma \approx 1$) while distant points approach independence ($\sigma \rightarrow 0$).
This implies that the GP is smooth, i.e.\ nearby input points produce similar outputs.

Just as we can sample random numbers from a distribution, we can draw random functions from a GP.
We choose a set of test points $X_*$ (the reason for the subscript $*$ will become clear in a moment), calculate the covariance matrix $\Sigma = \sigma(X_*, X_*)$, and generate multivariate normal samples from $\N(\zero, \Sigma)$.
We can then plot the input-output points as smooth curves, as in the top panel of Fig.~\ref{fig:gp}.

Of course, simply generating random functions is not particularly useful---we want to use a GP to interpolate a computer model.
Suppose we have a model which takes a vector of input parameters $\x$ and produces an output $y$ according to some unknown GP $f(\x)$; for example, $f$ could be a hydrodynamic model with input parameters $\x = (\tau_0, \eta/s)$ and the output could be elliptic flow $v_2$.
We choose a set of training points $X$, run the model at each point, and observe a set of outputs $\y$.
Now, instead of completely random functions, we desire functions which pass through (interpolate) all the training points $(X, \y)$.
This is achieved by \emph{conditioning} the GP on the training data to yield a predictive distribution for $y$ at any input point $\x$.
Recalling the test points $X_*$, the predictive distribution for the corresponding outputs $\y_*$ is the multivariate normal distribution
\begin{equation}
  \begin{aligned}
    \y_* &\sim \N(\muvec, \Sigma), \\
    \muvec &= \sigma(X_*, X)\sigma(X, X)^{-1}\y, \\
    \Sigma &= \sigma(X_*,X_*) - \sigma(X_*,X)\sigma(X,X)^{-1}\sigma(X,X_*).
  \end{aligned}
  \label{eq:cond}
\end{equation}
See the bottom panel of Fig.~\ref{fig:gp} for an example of conditioning a GP on one-dimensional training data.

We emphasize that the prediction $\y_*$ is not constant, but a \emph{probability distribution} for the model outputs at $X_*$.
As demonstrated in Fig.~\ref{fig:gp}, the predictive distribution is narrow when near the training points and wide when far away, hence, it reflects the true state of knowledge of the interpolation.
This is accomplished without assuming a parametric form for the model---we must only assume that the model is a GP with a specified covariance function.

\colfig[t]{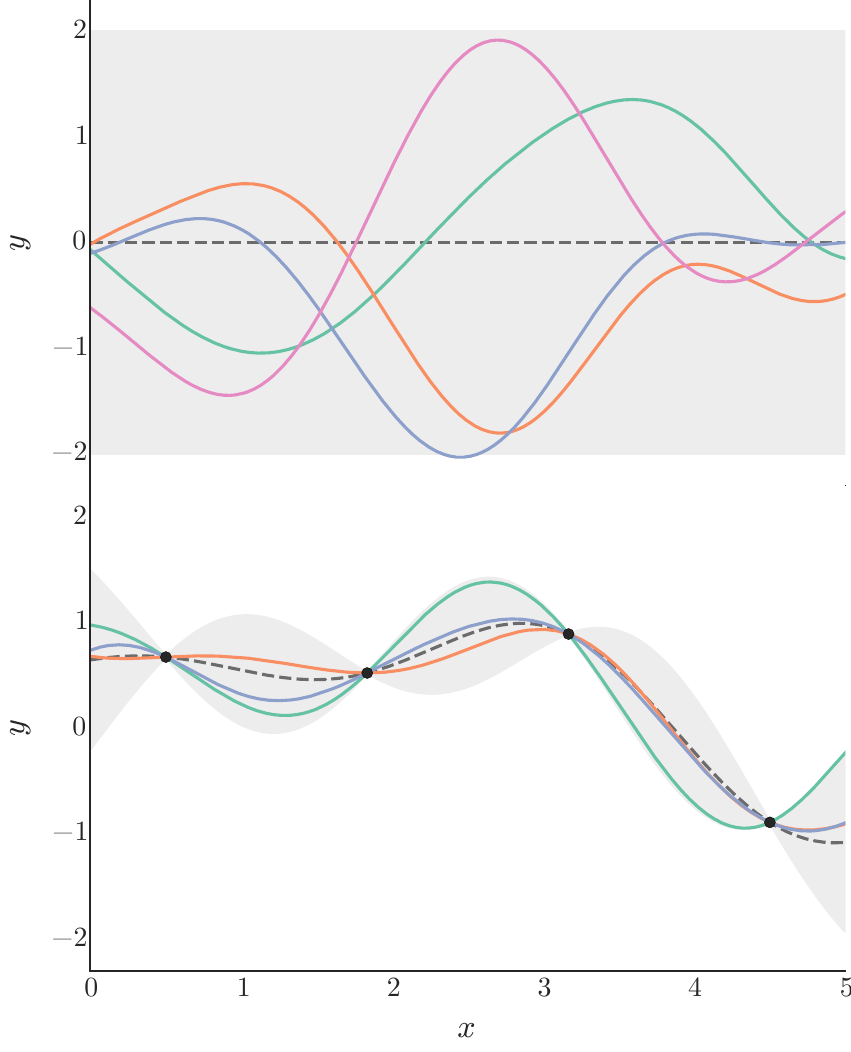}{
  Top: Random functions drawn from a Gaussian process using a squared-exponential covariance function with length scale $\ell = 1$.
  Bottom: Functions drawn from a GP conditioned on the training points indicated by dots.
  In both plots, the dashed line represents the GP mean and the grey band is twice the GP standard deviation (roughly 95\% confidence interval).
}

\subsection{Computer experiment design}

The full event-by-event model is to be evaluated on a set of $m$ training points $X = \{\x_1, \ldots, \x_m\}$, where each $\x_i$ is an $n$-dimensional vector of input parameters, so $X$ may be viewed as an $m \times n$ design matrix.
This subsection details the choice of input parameters and their distribution in parameter space.

For the present study, we choose a set of $n = 5$ input parameters
\begin{equation}
  \x = (\text{Norm}, \text{I.C.\ param}, \tau_0, \eta/s, \tau_\pi)
\end{equation}
where
\begin{itemize}
  \item Norm is the overall normalization factor, a multiplicative constant that determines how much entropy is deposited in the initial condition.
  \item I.C.\ param is a parameter specific to each initial condition model.
    For the Glauber model the parameter is $\alpha$, which controls how entropy is distributed to wounded nucleons and binary collisions;
    for the KLN model it is $\lambda$, a dimensionless exponent in the saturation scale parametrization.
    Both are related to the centrality dependence of multiplicity.
  \item $\tau_0$ is the QGP thermalization time and the starting time for hydrodynamic evolution.
  \item $\eta/s$ is the shear viscosity to entropy density ratio of the QGP, assumed to be fixed throughout the hydrodynamic evolution stage.
  \item $\tau_\pi$ is the shear stress relaxation time, which dictates how quickly the hydro medium relaxes to the Navier-Stokes limit.
    Since the relaxation time is a function of the shear viscosity and temperature and thus cannot be tuned explicitly, we use the coefficient $k_\pi$ in the relation $\tau_\pi = 6k_\pi\eta/(sT)$ as a tunable parameter.
\end{itemize}
We set intentionally large ranges for each parameter, summarized in Table~\ref{tab:design}.
In this work, we fix several auxiliary parameters to reasonable defaults:
nucleons are assumed to be disks with size determined by the inelastic nucleon-nucleon cross section $\sigma_\text{NN}$, and the hydro to micro switching temperature is set to 165 MeV, just below the equation of state transition temperature.
However, the method can handle arbitrary numbers of parameters provided the sample size is sufficiently large.

\begin{table}[t]
  \caption{
    \label{tab:design}
    Input parameter ranges for the Glauber (Glb) and KLN initial condition models and for the hydrodynamic model.
  }
  \begin{ruledtabular}
  \begin{tabular}{lll}
    Parameter & Description & Range \\
    \paddedhline
    Glb Norm & Overall normalization & 20--60 \\
    Glb $\alpha$ & Wounded nucleon / binary coll. & 0.05--0.30 \\
    KLN Norm & Overall normalization & 5--15 \\
    KLN $\lambda$ & Saturation scale exponent & 0.1--0.3 \\
    $\tau_0$ & Thermalization time & 0.2--1.0 fm \\
    $\eta/s$ & Specific shear viscosity & 0--0.3 \\
    $k_\pi$ & Shear relaxation time coeff. & 0.2--1.1 \\
  \end{tabular}
  \end{ruledtabular}
\end{table}

The training points $X = \{\x_1, \ldots, \x_m\}$ must be chosen to simultaneously optimize emulator accuracy and computation time.
Perhaps the most obvious design strategy is a uniform grid (factorial design), e.g.\ $k$ evenly-spaced points in each dimension.
Unfortunately, this leads to a total sample size $m = k^n$ which even for a modest $k = 10$ and $n = 5$ is intractably large.

A popular algorithm for generating efficient design points is maximin Latin hypercube sampling \cite{Morris:1995lh}.
This method produces space-filling randomized designs with several desirable properties:
\begin{itemize}
  \item The \emph{minimum} distance between points is \emph{maximized}, thus avoiding large gaps and tight clusters.
  \item Projections of the design into lower dimensions are uniformly distributed.
\end{itemize}
Figure~\ref{fig:design} illustrates these traits.
A Latin hypercube design with a relatively small sample size provides an efficient scaffolding of parameter space for interpolation by a GP emulator.
As a general rule of thumb, a sample size $m \sim 10n$ yields acceptable interpolation accuracy \cite{Loeppky:2009ss} and is a common choice for an initial experiment with limited computation time, however there is no harm in a larger sample.

We use a 256 point Latin hypercube design across the $n = 5$ input parameters; Fig.~\ref{fig:design} shows a two-dimensional projection.
At each design point, we have executed the event-by-event model \order 3 times in each of six centrality bins 0--5\%, 10--15\%, \ldots, 50-55\%, for both the Glauber and KLN models, yielding \order 7 events in total.
Two design points that were very near to the edge of the design space gave non-physical results and have been discarded, so the operational design has $m = 254$ points.

Figure~\ref{fig:prior_draws} shows the postprocessed observables $\avg\nch$, $\vnk 2 2$, $\vnk 3 2$ as a function of centrality for each point in the design.
The results have a broad distribution which is a direct result of the wide ranges of input parameters.
There is some statistical error present in $\vnk 3 2$ due to insufficient quantities of events.

Note that these results constitute the training data for the GP emulator, not any kind of best-fit.

\colfig[t]{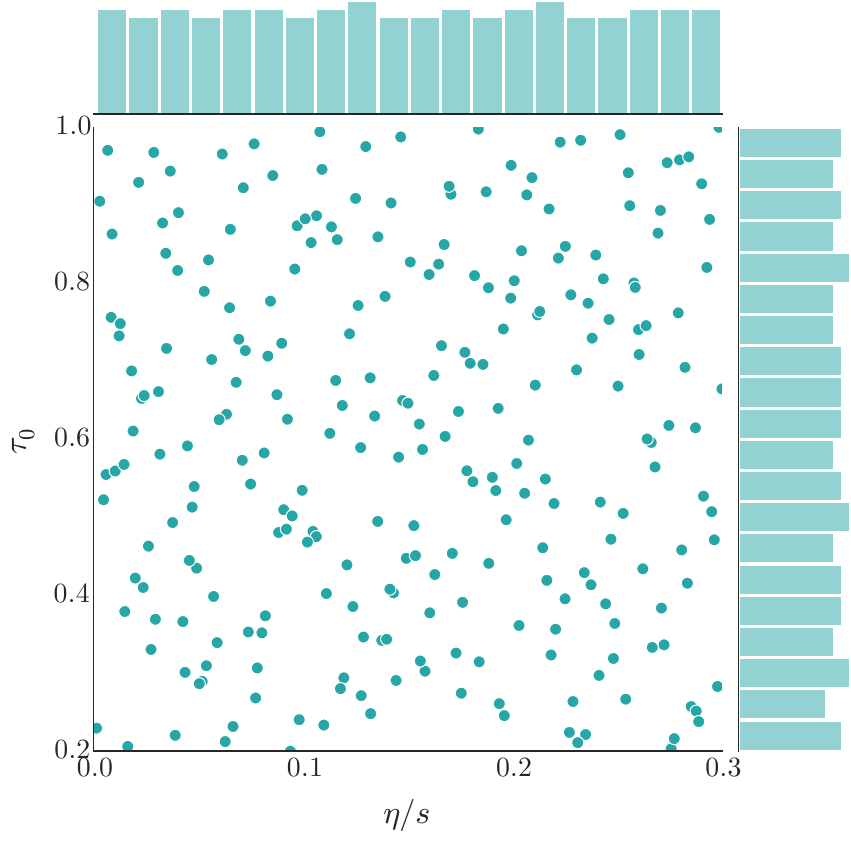}{
  The Latin hypercube experiment design projected into the $(\eta/s, \tau_0)$ dimensions.
  All other parameters also vary across the design, so the points that appear very close in the projection are not necessarily close in the full-dimensional space.
  The edge histograms show the distributions flattened into one dimension; note that they are space-filling and approximately uniform.
}

\subsection{\label{sec:multiout}Multivariate output}

\widefig{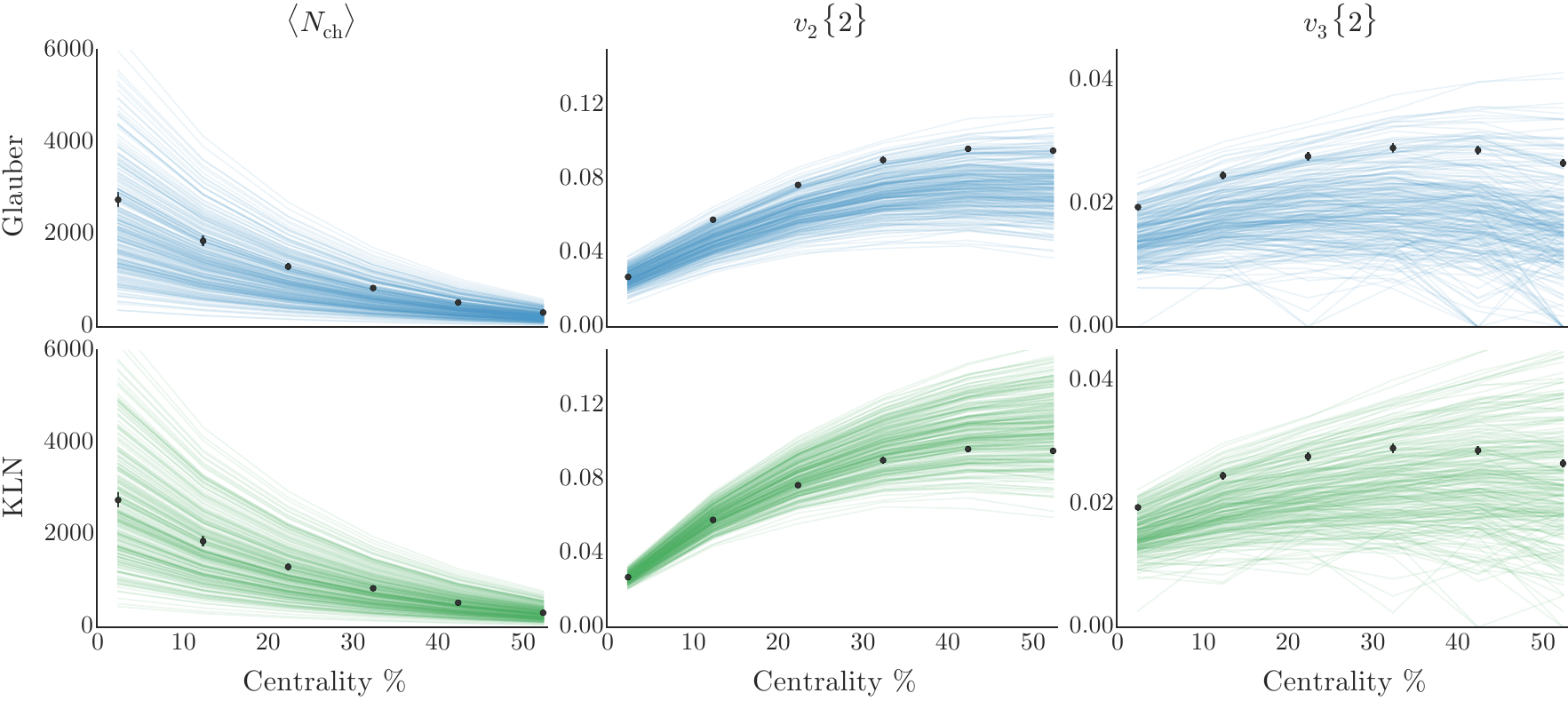}{
  Model calculations from Glauber (top, blue) and KLN (bottom, green) initial conditions.
  Each plot has 254 lines corresponding to the 254 Latin hypercube design points.
  From left to right:
  average charged-particle multiplicity $\avg\nch$,
  elliptic flow two-particle cumulant $\vnk 2 2$,
  and triangular flow two-particle cumulant $\vnk 3 2$.
  Data points are experimental measurements from ALICE \cite{Abelev:2014mda}.
}

\colfig[b!]{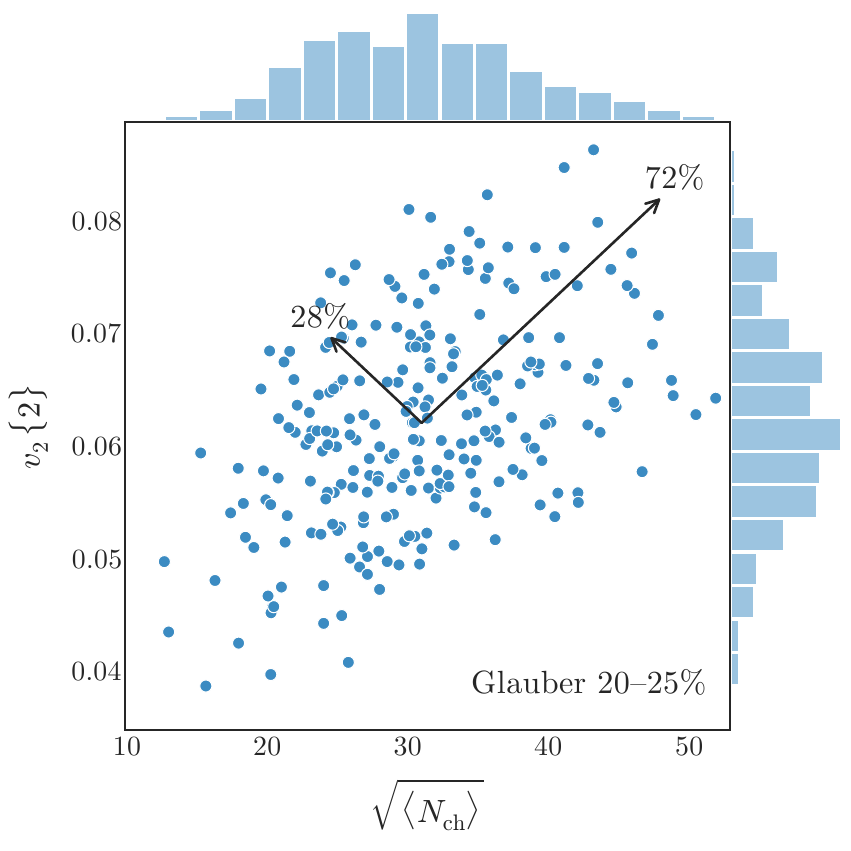}{
  Principal component decomposition of the observables $\sqrt{\avg\nch}$, $\vnk 2 2$ for the Glauber model in 20--25\% centrality.
  Each data point represents a model calculation and the edge histograms show the approximate normal distribution of each observable.
  Arrows represent the PC vectors with lengths proportional to the explained variance.
}

Gaussian processes are fundamentally scalar functions, but computer models often produce multivariate output.
In general, the model takes the $m \times n$ design matrix $X$ and computes an $m \times p$ output matrix $Y$.
The present event-by-event model has $p = 18$ outputs (three observables each in six centrality bins).

An obvious workaround is to use independent GP emulators for each of the $p$ outputs, however, this would neglect correlations and quickly become unwieldy for higher-dimensional output.
Instead, we decompose the outputs into orthogonal linear combinations called principal components (PCs) and emulate each transformed PC.
The PCs are uncorrelated by construction and can also be used to reduce the dimensionality of the output space.
Figure~\ref{fig:pc_scatter} shows an example PC decomposition.

To calculate the PCs, we first subtract the mean of the output data $Y$ so that each column has mean zero, then compute the eigendecomposition of the sample covariance matrix $Y\tran Y$:
\begin{equation}
  Y\tran Y = U \Lambda U\tran,
  \label{eq:cov}
\end{equation}
where $U$ is an orthogonal $p \times p$ matrix containing the eigenvectors of $Y\tran Y$ and $\Lambda$ is diagonal containing the eigenvalues $\lambda_1, \ldots, \lambda_p$ in non-increasing order.
$U$ now defines a linear transformation which ``rotates'' the output data $Y$ into PC space:
\begin{equation}
  Z = \sqrt m \, YU,
\end{equation}
where $Z$ is an $m \times p$ matrix (same shape as $Y$) of the transformed PCs.
The eigenvalues $\lambda_i$ represent the variance explained by principal component $i$; since they are sorted in non-increasing order, the \emph{fraction} of the variance explained by the first $q \leq p$ PCs is
\begin{equation}
  V(q) = \frac{\sum_{i=1}^q \lambda_i}{\sum_{i=1}^p \lambda_i}.
\end{equation}
Often, the first few PCs describe most of the variance, as demonstrated for the present data in Fig.~\ref{fig:pc_var}.
Hence we can construct a reduced-dimension transformation with minimal loss of precision by choosing $q < p$ so that $V(q)$ satisfies some threshold (e.g.~$V(q) \geq 0.99$) and taking only the first $q$ columns of $U$:
\begin{equation}
  Z_q = \sqrt m \, YU_q,
\end{equation}
where $Z_q$ is now an $m \times q$ matrix.

We may now use $q$ independent GP emulators for each of the columns of $Z_q$.
GPs are conditioned on the design $X$ according to Eq.~\eqref{eq:cond} and predict the PCs $Z_*$ at arbitrary test points $X_*$ which are then transformed back to physical space as
\begin{equation}
  Y_* = \frac{1}{\sqrt m} Z_* U\tran.
\end{equation}

There is an important caveat for principal components:
the original data $Y$ must have a multivariate normal distribution for the transformed PCs $Z$ to be uncorrelated.
There is no guarantee that a particular model will produce normally-distributed outputs so this must be verified on a case-by-case basis.
For the present event-by-event model we perform the following steps:
\begin{enumerate}
  \item Assess the normality of each observable $\avg\nch$, $\vnk 2 2$, $\vnk 3 2$.
    While the flow cumulants are approximately normal without modification, we take the square root of multiplicity $\sqrt{\avg\nch}$ to obtain a normal distribution, as shown in Fig.~\ref{fig:pc_scatter}.
  \item Divide each observable by its corresponding experimental value from ALICE \cite{Abelev:2014mda}.
    This converts everything to unitless quantities of order one.
  \item Multiply each observable by a manually specified weight factor, ratios 1.2~:~1.0~:~0.6 for observables $\avg\nch$~:~$\vnk 2 2$~:~$\vnk 3 2$.
    These subjective weights encourage the model to fit more strongly to more fundamental observables, e.g.\ we prefer a model that describes $\nch$ and $v_2$ at the expense of $v_3$ rather than fitting $v_3$ with incorrect $\nch$.
    The weights will be discussed further in the results, Sec.~\ref{sec:results}.
  \item Concatenate the unitless, weighted data into a $254 \times 18$ matrix $Y$, where each row corresponds to a design point and each column to an observable and centrality bin.
  \item Subtract the mean of each column and transform $Y$ into principal components $Z_q$ with $q = 5$ PCs, retaining over 99\% of the variance as shown in Fig.~\ref{fig:pc_var}.
    The PC transformation matrices $U_q$, shown in Fig.~\ref{fig:pc_matrix}, reflect the natural correlations among observables, for example all observables are correlated in the first PC (${\sim}$75\% of total variance), while $N_\text{ch}$ is anti-correlated with the $v_n$ in the second PC (${\sim}$20\% of total variance).
\end{enumerate}
We invert these steps to transform PCs back to physical space.

In practice, the covariance method for computing principal components is prone to numerical error, so a more robust algorithm using the singular value decomposition (SVD) is preferred.
The SVD of the data $Y$ is
\begin{equation}
  Y = VDW\tran
  \label{eq:svd}
\end{equation}
where $V$, $W$ are orthogonal matrices containing the so-called left- and right-singular vectors of $Y$ and $D$ is diagonal containing the singular values.
Inserting \eqref{eq:svd} into \eqref{eq:cov} yields
\begin{equation}
  Y\tran Y = W D^2 W\tran = U \Lambda U\tran,
\end{equation}
hence the singular values $D$ are the square root of the eigenvalues $\Lambda$ and the right singular vectors $W$ are the eigenvectors $U$.

\colfig[t]{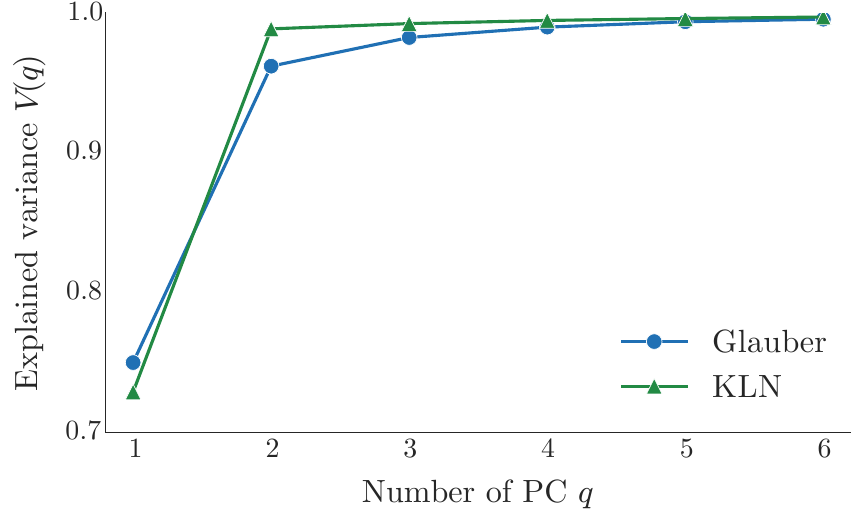}{
  Fraction of the variance $V(q)$ explained by the first $q$ principal components for Glauber (blue circles) and KLN (green triangles).
  $q = 5$ explains approximately 99\% of the total variance, a significant reduction from the original 18 dimensions.
}

\colfig[t]{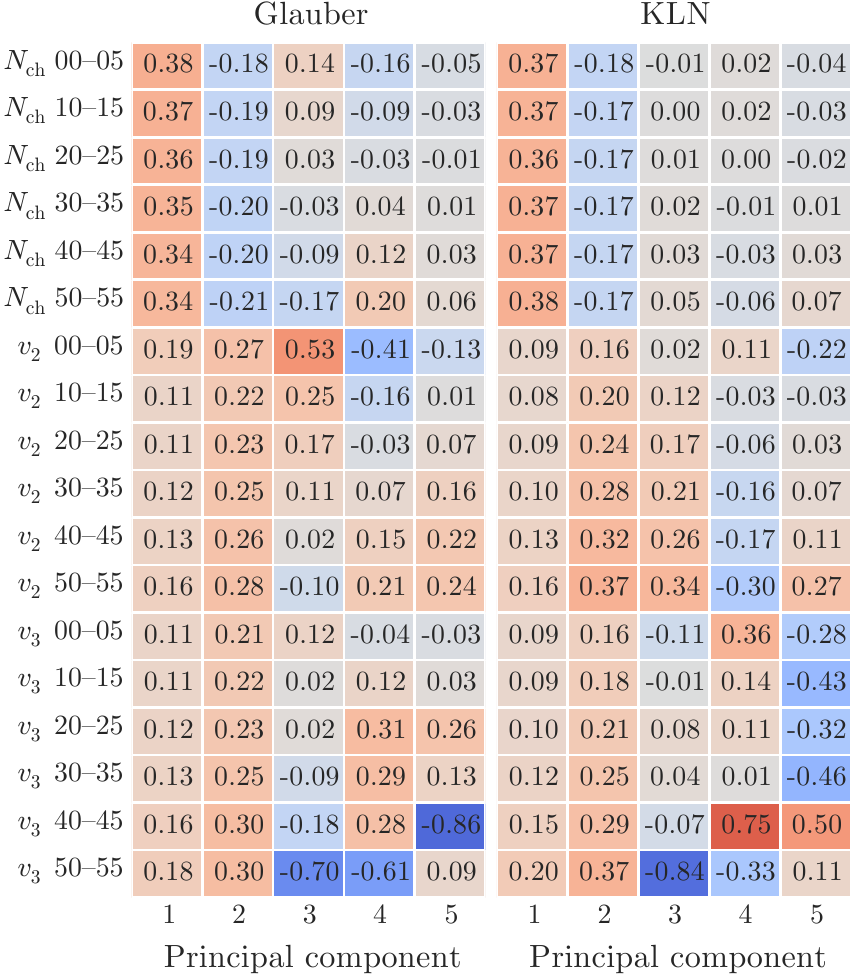}{
  Visualization of the principal component transformation matrices $U_q$ for Glauber (left) and KLN (right).
  The numerical values of each matrix element are annotated and color-coded, where darker red indicates more positive values, darker blue indicates more negative, and grey indicates zero.
}

\subsection{Constructing and validating the emulator}

\widefig{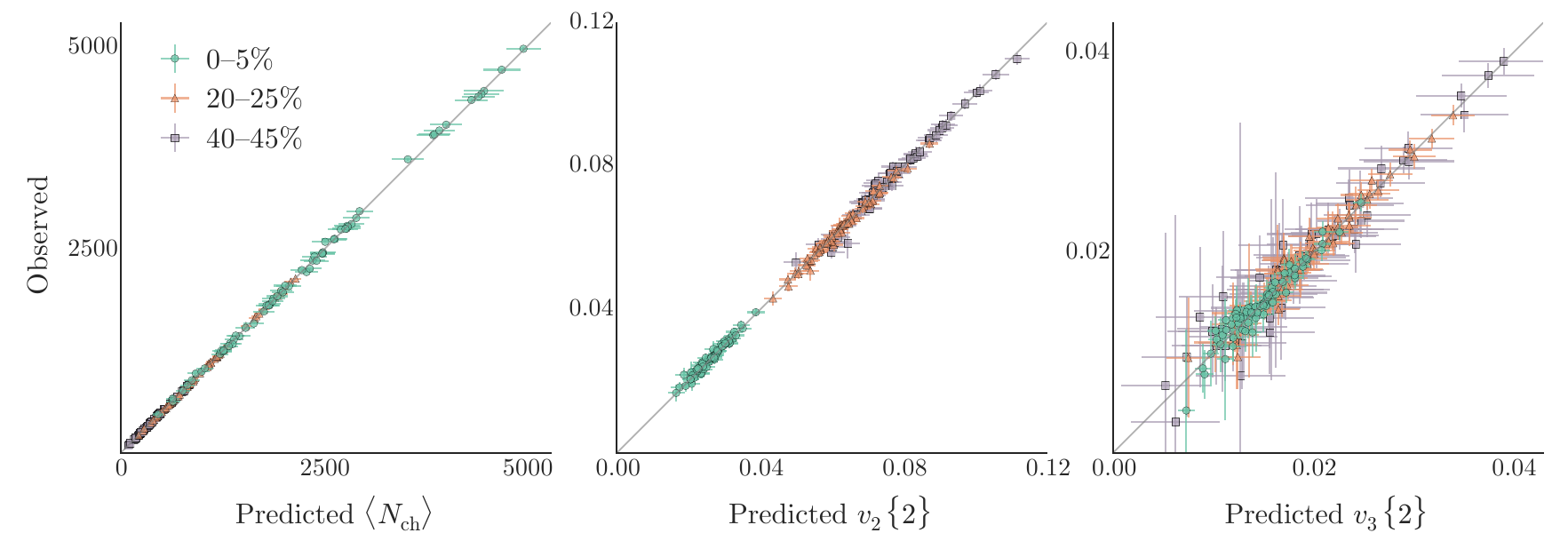}{
  Validation of the Gaussian process emulator for the Glauber model.
  Each plot shows emulator predictions against explicit calculations for the 64 validation design points in centrality bins 0--5\% (green circles), 20--25\% (orange triangles), and 40--45\% (purple squares).
  The $x$-value of each data point is the emulator prediction with $2\sigma$ (95\%) horizontal error bars, the $y$-value is the explicit calculation with $2\sigma$ (95\%) vertical error bars, and the diagonal grey line represents $y = x$.
}

We emulate the model by conditioning independent Gaussian processes on each of the principal components $Z_q$ and the input design $X$ according to Eq.~\eqref{eq:cond}.
Model outputs inevitably include statistical noise, i.e.\ we cannot compute $y = f(\x)$ exactly, only $y = f(\x) + \epsilon$ where $\epsilon$ is Gaussian noise.
This is accounted for by adding a noise term to the diagonal of the covariance matrix:
\begin{equation*}
  \sigma(\x, \x') \rightarrow \sigma(\x, \x') + \sigma^2_n\delta_{\x\x'},
\end{equation*}
where $\sigma^2_n$ is the variance of the noise and $\delta_{\x\x'}$ is a Kronecker delta.
Effectively, the noise term relaxes the requirement that the GP must pass exactly through each training point.

We use a squared-exponential covariance function with a noise term:
\begin{equation}
  \sigma(\x, \x') = \sigma_\text{GP}^2 \exp\Biggl[ -\sum_{k=1}^n \frac{(x_k - x'_k)^2}{2\ell_k^2} \Biggr] + \sigma_n^2\delta_{\x\x'},
  \label{eq:full_cov}
\end{equation}
where $\sigma_\text{GP}^2$ is the overall variance of the GP and $\ell_k$ is the characteristic length scale for dimension $k$.
These \emph{hyperparameters} $(\sigma_\text{GP}$, $\sigma_n$, $\ell_k)$ are not known a priori and must be estimated from the training data, however, in the present case predictions appear to be relatively insensitive to the precise choice of hyperparameters, as will be demonstrated promptly.
For details about the selection of hyperparameters see the \hyperref[app:train]{Appendix}.

As with any interpolation scheme, the GP emulator must be validated to ensure it faithfully predicts model output.
In other words, given an arbitrary test point $\x_*$, the GP prediction at $\x_*$ should agree (within its uncertainty) with an explicit computation at $\x_*$.
To this end, we have generated a separate 64-point Latin hypercube validation design $X_*$, evaluated the full event-by-event model at each validation point just as for the training design $X$, and predicted the model outputs at $X_*$ using the GP emulator.

Figure~\ref{fig:validation} validates that the emulator does indeed faithfully predict the model.
Recall that the emulator provides probability distributions of finite width, so it need not predict every validation point exactly---in fact, in the ideal case the residuals would have a normal distribution with mean zero.
Most of the uncertainty visible in Fig.~\ref{fig:validation} is actually due to the statistical noise in the flow cumulants, especially $\vnk 3 2$.
The emulator accurately accounts for the noise present in the underlying data.

\section{Calibration}

\newcommand{\z}{\mathbf z}
\newcommand{\xs}{\x_\star}
\newcommand{\zs}{\z_\star}
\newcommand{\yexp}{\y_\text{exp}}
\newcommand{\zexp}{\z_\text{exp}}

With the validated Gaussian process emulator in hand, it may be used as a fast surrogate to the full event-by-event model for calibration.
\emph{Calibration} means tuning the model input parameters so that the output optimally agrees with experimental data and in the process extracting probability distributions for each parameter.
Recall the input parameters are
\begin{equation*}
  \x = (\text{Norm}, \text{I.C.\ param}, \tau_0, \eta/s, k_\pi).
\end{equation*}
Presumably there exists a \emph{true} set of parameters $\xs$; the task now is to find the probability distribution of $\xs$ given the training data $(X, Y)$ and experimental measurements $\yexp$.
This distribution may be framed in terms of Bayes' theorem as
\begin{equation}
  P(\xs|X,Y,\yexp) \propto P(X,Y,\yexp|\xs) P(\xs)
\end{equation}
where
\begin{itemize}
  \item $P(\xs)$ is the \emph{prior} probability which embodies initial knowledge of $\xs$;
  \item $P(X,Y,\yexp|\xs)$ is the likelihood:
    the probability of observing $(X, Y, \yexp)$ given a proposed value of $\xs$; and
  \item $P(\xs|X,Y,\yexp)$ is the \emph{posterior} probability for $\xs$ given the observations $(X, Y, \yexp)$.
    This is the probability distribution we wish to construct.
\end{itemize}
In general Bayes' theorem has a normalization constant which has been omitted since we are only concerned with relative probabilities.

The remainder of this section applies the methodology from \cite{OHagan:2006ba,Higdon:2008cmc,Higdon:2014tva} to calibrate the model and determine the posterior probability for $\xs$.

\subsection{MCMC}

\widefig{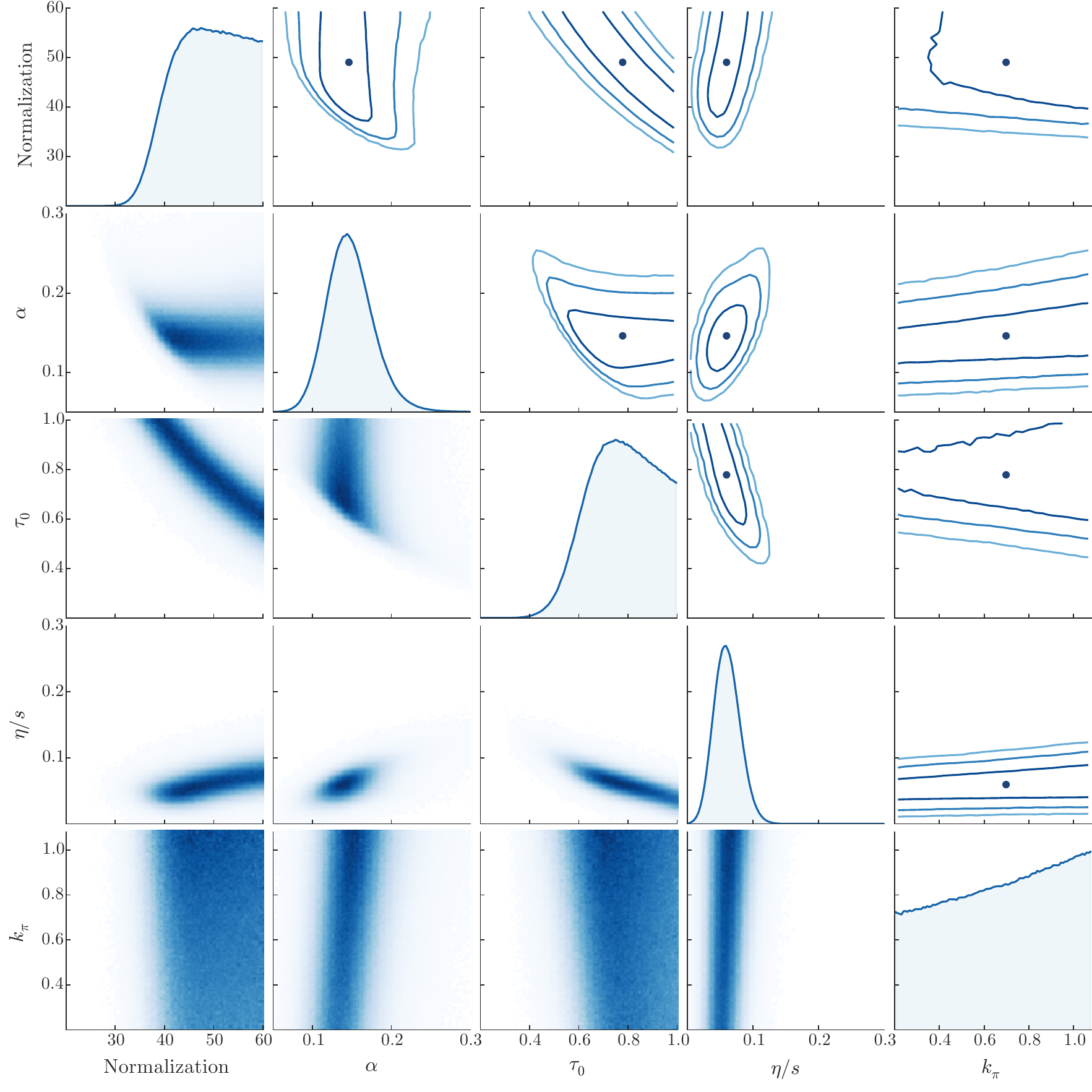}{
  Posterior marginal and joint distributions of the calibration parameters for the Glauber model.
  On the diagonal are histograms of MCMC samples for the respective parameters,
  on the lower triangle are two-dimensional scatter histograms of MCMC samples showing the correlation between pairs of parameters,
  and on the upper triangle are approximate contours for 68\%, 95\%, and 99\% confidence regions along with a dot indicating the median.
}

The workhorse of any Bayesian calibration is Markov chain Monte Carlo (MCMC), a powerful and flexible method for directly sampling the posterior probability.
Perhaps the most common version is the Metropolis-Hastings algorithm, which generates a random walk through parameter space by accepting or rejecting steps based on the posterior probability.
For a large number of steps the samples of the random walk equilibrate to the posterior distribution.
We use the affine-invariant ensemble sampler for MCMC \cite{Goodman:2010en,FM:2013mc}, an alternative algorithm that uses a large ensemble of interdependent walkers.
Ensemble sampling notably has a much shorter autocorrelation time than Metropolis-Hastings sampling and hence converges more quickly to the equilibrium distribution.

\widefig{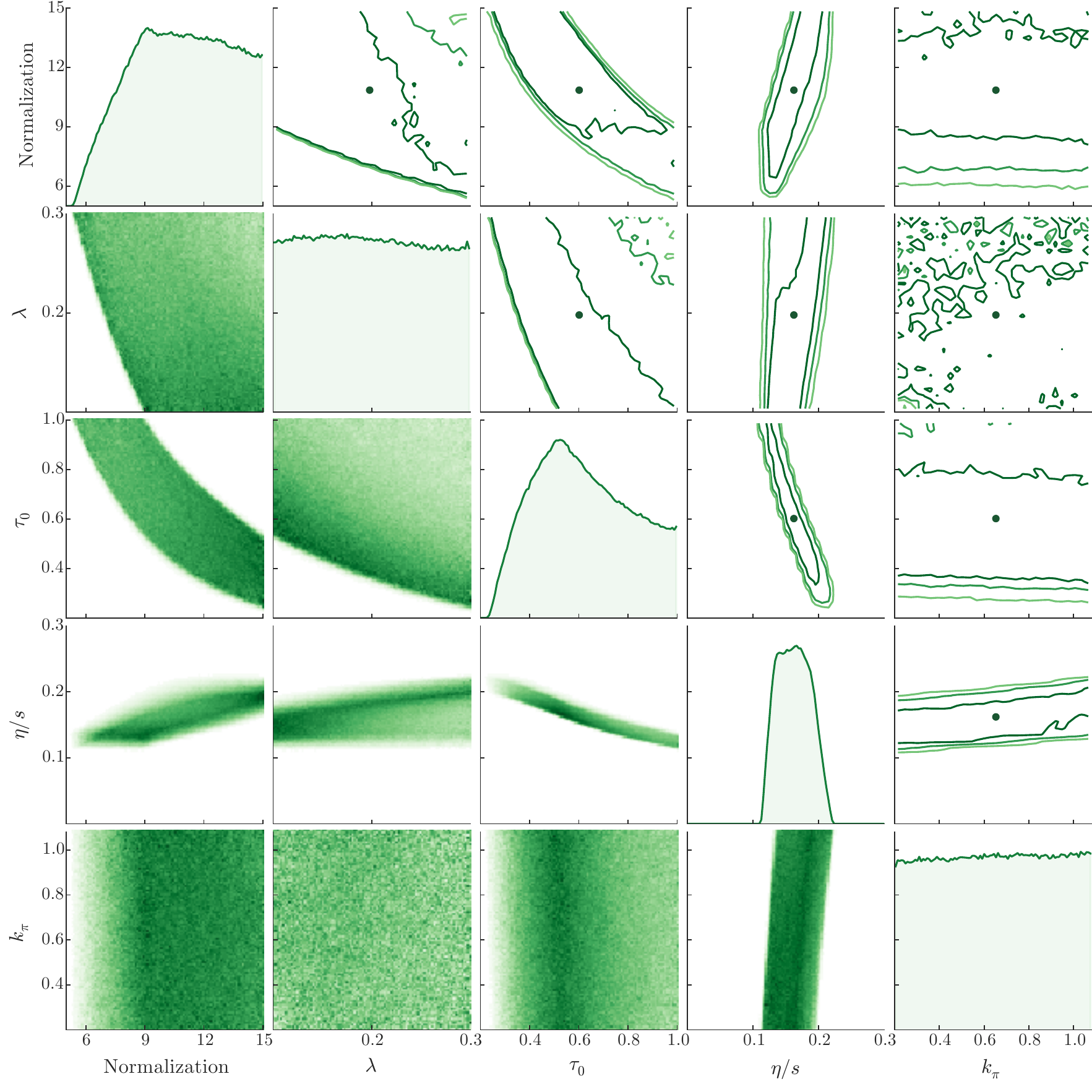}{
  Same as Fig.~\ref{fig:cal_post_glb} for the KLN model.
}

The MCMC algorithm samples proposal points $\xs$ and calculates the posterior probability at each point via Bayes' theorem.
We place a non-informative flat prior on $\xs$, that is, the prior probability is constant within the design range (Table~\ref{tab:design}) and zero outside.
We evaluate the likelihood in principal component space:
\begin{equation}
  P(\zexp|\xs) \propto \exp\biggl\{ -\frac{1}{2} (\zs - \zexp)\tran \Sigma_z^{-1} (\zs - \zexp) \biggr\},
  \label{eq:likelihood}
\end{equation}
where $\zexp$ is the PC transform of the experimental data, $\zs$ is the emulator prediction of the PCs at $\xs$, and $\Sigma_z$ is the covariance (uncertainty) matrix on the PCs assuming normally-distributed errors.
Given the flat prior, the posterior $P(\xs|\zexp)$ is equal to the likelihood within the design range and zero outside.

There are a number of sources of uncertainty including experimental statistical and systematic error, model statistical and systematic error, and emulator uncertainty.
In the present study, we do not attempt to precisely account for each contribution, for this would inevitably require dubious assumptions about systematic error correlations and the unknown error of the model.
We assume a simple fractional error on the principal components, i.e.\ the covariance matrix is
\begin{equation}
  \Sigma_z = \text{diag}(\sigma^2_z\,\zexp),
\end{equation}
where $\sigma^2_z$ is a manually set constant, $\sigma_z = 0.06$ for the present study, to account for the typical experimental error of 3--5\% \cite{Abelev:2014mda} plus some additional uncertainty.
While this is itself a rough assumption, it is perhaps no worse than the alternative, since experimental systematic errors are typically estimated percentages themselves and the principal component transformation automatically includes natural correlations among observables.
The primary goal of this study is to develop and test a model-to-data comparison framework; details such as the precise treatment of uncertainties can be improved later.

\widefig{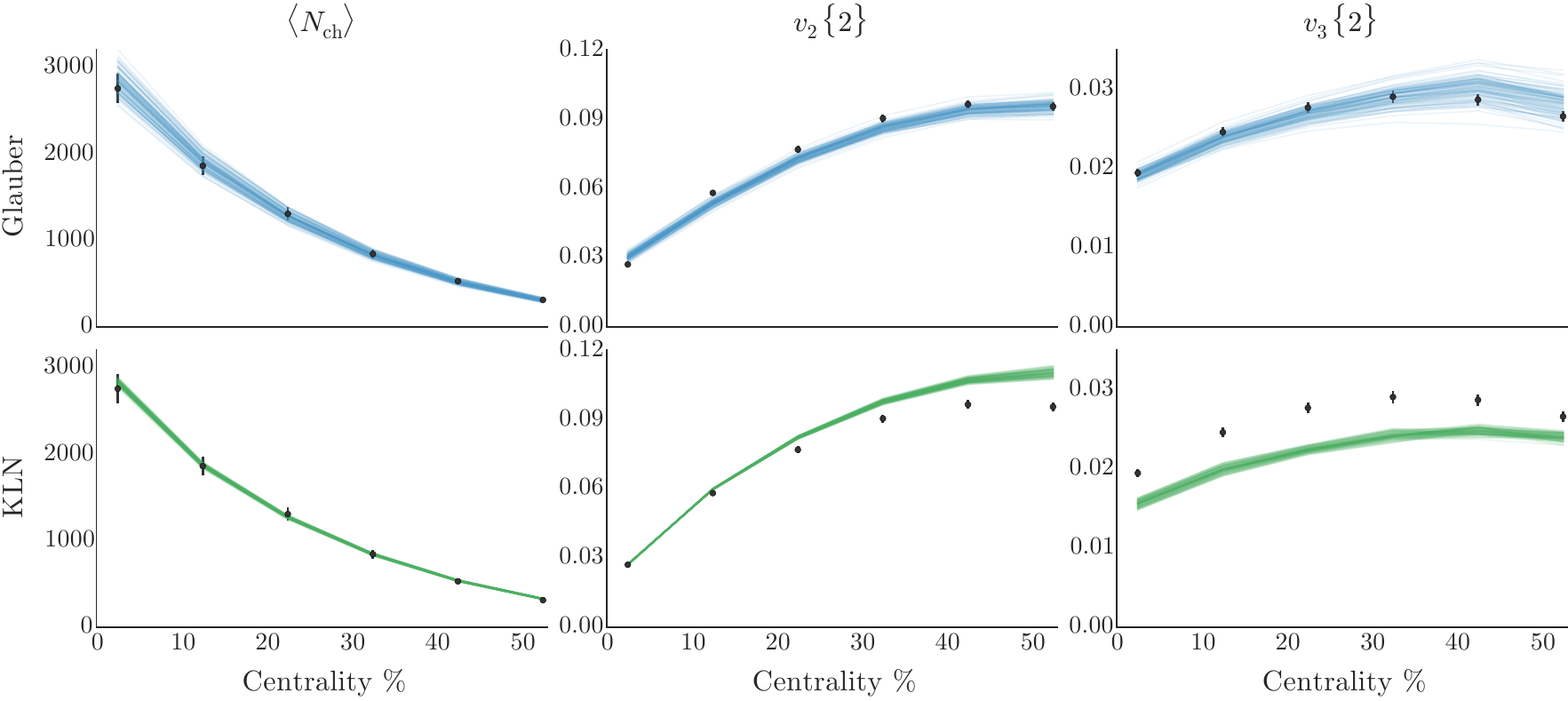}{
  Random realizations of the calibrated posterior for Glauber (top, blue) and KLN (bottom, green) initial conditions.
  Similar to Fig.~\ref{fig:prior_draws} except the lines are posterior emulator predictions instead of explicit prior calculations.
}

We run \order 6 MCMC steps to allow the chain to equilibrate, discard these ``burn-in'' samples, then run \order 7 steps to generate the posterior distribution.

\vfill

\subsection{\label{sec:results}Results}

The primary MCMC calibration results are presented in Figs.~\ref{fig:cal_post_glb} and \ref{fig:cal_post_kln} for the Glauber and KLN models, respectively.
These are visualizations of the posterior probability distributions of the true parameters $\xs$, including the distribution of each individual parameter and all correlations.
The diagonal histograms show the marginal distributions for each parameter (all other parameters integrated out);
the lower-triangle plots are two-dimensional scatter histograms of joint distributions between pairs of parameters, where darker color denotes higher probability density;
and the upper triangle has contour plots of the same joint distributions, where the contour lines enclose the 68\%, 95\%, and 99\% confidence regions.

A wealth of information may be gained from these posterior visualizations; the following highlights some important features.

Focusing on the Glauber results in Fig.~\ref{fig:cal_post_glb}, we see the shear viscosity $\eta/s$ (fourth diagonal plot) has a narrow approximately normal distribution located near the commonly quoted value 0.08.
As expected, $\eta/s$ is tightly constrained by experimental flow data.
Going across the fourth row, we observe nontrivial correlations among $\eta/s$ and other parameters, for example, $\eta/s$ and the hydrodynamic thermalization time $\tau_0$ are negatively correlated (fourth row, third column).
As $\tau_0$ increases, the medium expands as a fluid for less time, so less flow develops, and viscosity must decrease to compensate.

Both $\tau_0$ and normalization (third and first diagonals) have broad distributions without strong peaks, and they are strongly-correlated (third row, first column).
This is because the hydrodynamic model is boost-invariant and lacks any pre-equilibrium dynamics, so $\tau_0$ is effectively an inverse normalization factor.
The joint distribution shows a narrow acceptable band whose shape is governed by the inverse relationship.

The wounded nucleon / binary collision parameter $\alpha$ (second diagonal) has a roughly-normal distribution located near the typical value 0.12.
It is mainly related to the slope of multiplicity vs.\ centrality and hence has a nontrivial correlation with normalization and $\tau_0$, e.g.\ we can decrease the normalization to the lower end of its distribution provided we also increase $\alpha$ to compensate.

Meanwhile, the shear stress relaxation time coefficient $k_\pi$ (fifth diagonal) has an almost flat distribution and its joint distributions show no correlations.
Evidently, this parameter does not influence flow coefficients or multiplicity.

The KLN results in Fig.~\ref{fig:cal_post_kln} generally exhibit wider, less normal distributions than Glauber.
This suggests that KLN is somewhat less flexible than Glauber, so its overall behavior is relatively insensitive to the specific values of input parameters.

The shear viscosity $\eta/s$ has a narrow, irregular distribution covering the common value 0.20.
As with Glauber, $\eta/s$ has a negative correlation with $\tau_0$, there is a strong inverse relationship between normalization and $\tau_0$, and $k_\pi$ has no effect.
The KLN parameter $\lambda$ has a flat marginal distribution, but there are strongly excluded regions in the joint distributions with normalization and $\tau_0$.
This appears to be the same effect as observed with Glauber $\alpha$, except the dependence on $\lambda$ is significantly weaker.

The posteriors may be validated by drawing samples from the calibrated distributions and visualizing the corresponding emulator predictions:
if the model is correct and properly calibrated, the posterior samples will be close to experimental measurements.
Figure~\ref{fig:post_draws} confirms---for the most part---that the posteriors are indeed tightly clustered around the data points.
Visualizations such as this will always have some uncertainty since samples are drawn from the full posterior, however, Fig.~\ref{fig:post_draws} has markedly narrower clusters than Fig.~\ref{fig:prior_draws}, in which the input parameters varied across their full ranges and were not tuned to match experiment.

\begin{table*}
  \caption{
    \label{tab:posterior}
    Quantitative summary of posterior distributions.
    For each parameter, the previous estimate \cite{Shen:2011zc,Heinz:2011kt,Shen:2013pc}, mean, median, and confidence intervals are given.
  }
  \begin{ruledtabular}
\begin{tabular}{lllllccc}
  & Parameter & Prev.\ est. & Mean & Median & 68\% C.I. & 95\% C.I. & 99\% C.I. \\
  \paddedhline
  \multirow{5}{*}{\rotatebox{90}{Glauber\ \ }}
  & Norm. & 57 & 48.9 & 49.0 & 41.6--56.4 & 36.5--59.4 & 33.9--59.9 \\
  & $\alpha$ & 0.12 & 0.148 & 0.146 & 0.119--0.176 & 0.0954--0.212 & 0.0808--0.242 \\
  & $\tau_0$ & 0.6 & 0.776 & 0.778 & 0.638--0.922 & 0.527--0.987 & 0.461--0.997 \\
  & $\eta/s$ & 0.08 & 0.0604 & 0.0595 & 0.0407--0.0801 & 0.0244--0.101 & 0.0149--0.116 \\
  & $k_\pi$ & 0.5 & 0.682 & 0.698 & 0.373--0.978 & 0.228--1.08 & 0.206--1.09 \\
  \paddedhline
  \multirow{5}{*}{\rotatebox{90}{KLN\ \ }}
  & Norm. & 9.9 & 10.8 & 10.9 & 8.15--13.6 & 6.40--14.8 & 5.82--15.0 \\
  & $\lambda$ & 0.14 & 0.199 & 0.198 & 0.132--0.267 & 0.105--0.295 & 0.101--0.299 \\
  & $\tau_0$ & 0.6 & 0.620 & 0.602 & 0.415--0.846 & 0.302--0.975 & 0.265--0.995 \\
  & $\eta/s$ & 0.20 & 0.163 & 0.162 & 0.135--0.190 & 0.121--0.208 & 0.116--0.215 \\
  & $k_\pi$ & 0.5 & 0.651 & 0.653 & 0.347--0.955 & 0.223--1.07 & 0.205--1.09 \\
\end{tabular}
\end{ruledtabular}

\end{table*}

\colfig[b]{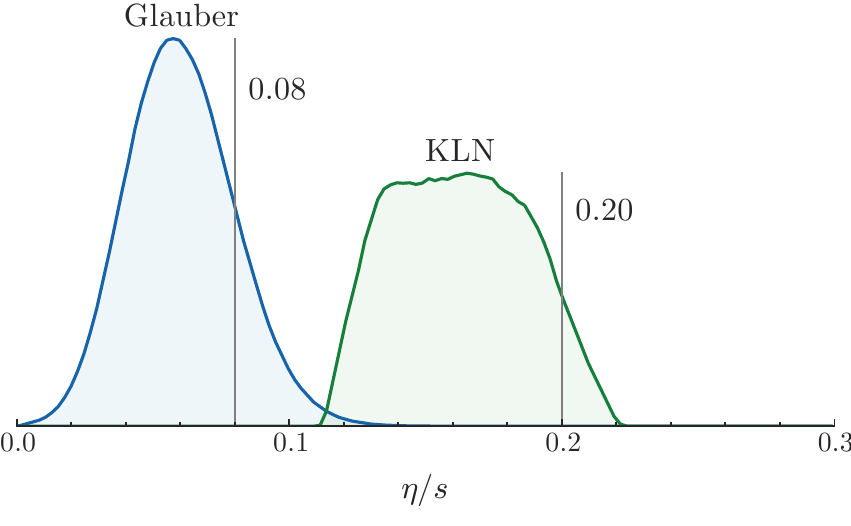}{
  Comparison of posterior distributions of $\eta/s$ for Glauber (blue) and KLN (green).
  These are the same histograms as in Figs.~\ref{fig:cal_post_glb} and \ref{fig:cal_post_kln}, expanded and placed on the same axis.
  The vertical grey lines indicate the common values 0.08 for Glauber and 0.20 for KLN \cite{Shen:2011zc,Heinz:2011kt}.
}

As shown in the top row of Fig.~\ref{fig:post_draws}, the Glauber model nearly fits the centrality dependence of all the present observables ($\avg\nch$, $\vnk 2 2$, $\vnk 3 2$).
The $v_3$ samples have a somewhat larger variance than the others, in part due to the underlying noise in the model calculations and also because $v_3$ is explicitly given a lower weight (recall that $\avg\nch$~:~$\vnk 2 2$~:~$\vnk 2 3$ are weighted 1.2~:~1.0~:~0.6).

The KLN results in the bottom row tell a somewhat different story, as they cannot fit all observables simultaneously.
While the fit to $\avg\nch$ is excellent, the ratio of $v_2$ to $v_3$ is simply too large and the model has no choice but to compromise between the two, similar to previous KLN results \cite{Qiu:2011hf}.
The posterior biases more towards $v_2$ than $v_3$ due to the explicit higher weight on $v_2$.

Figure~\ref{fig:post_compare} shows an expanded view of the $\eta/s$ marginal distributions for Glauber and KLN.
The Glauber distribution is approximately normal with mean ${\sim}$0.06 and 95\% confidence interval ${\sim}$0.02--0.10, consistent with but mostly below 0.08.
This is unsurprising and easily within the uncertainty of existing results.
KLN has a wider plateau-like distribution with mean ${\sim}$0.16 and 95\% confidence interval ${\sim}$0.12--0.21.
While the common estimate 0.20 was derived primarily from comparisons to $v_2$, the additional constraint from $v_3$ shifts the distribution to somewhat smaller values and causes the plateau shape:
rather than a strong peak, there is a range of values which all fit the data roughly equally.

Table~\ref{tab:posterior} quantitatively summarizes the posterior distributions for each parameter including basic statistics, confidence intervals, and comparisons to previous estimates from earlier work with the same models \cite{Shen:2011zc,Heinz:2011kt,Shen:2013pc}.
All previous estimates fall within 95\% confidence intervals, and most within 68\%.

\section{Conclusion}

We have applied modern Bayesian methodology to systematically compare an event-by-event heavy-ion collision model to experimental data.
We chose a set of salient model parameters including the shear viscosity $\eta/s$, evaluated the model over wide ranges of each parameter, and interpolated the results with a Gaussian process emulator.
Then, we used the emulator to calibrate the model to optimally reproduce experimental data and thereby extracted probability distributions for the true values of all model parameters simultaneously, including all correlations.

When properly calibrated, the Monte Carlo Glauber model provides a good simultaneous fit to experimental multiplicity and flow data, while the Monte Carlo KLN model fails to simultaneously fit elliptic and triangular flow.
The $\eta/s$ distributions for the Glauber and KLN models are consistent with the commonly quoted values 0.08 and 0.20, respectively, and in general the calibrated distributions reinforce and expand upon existing knowledge of these models.

This study represents a significant step forward in state-of-the-art model-to-data comparison and establishes a framework for future analysis.
Since the method does not reduce each parameter to a ``best-fit'' value but instead furnishes full probability distributions, it may be used to rigorously quantify uncertainties, examine correlations among parameters, and evaluate the efficacy of physical models, among other possibilities.
It is easily extensible to arbitrary numbers of parameters and physical observables and to different models.

Indeed, we plan to apply the methodology to a variety of other models, including the new initial condition model \trento---a flexible effective model which is ideal for this type of analysis \cite{Moreland:2014oya}---and a 3+1D viscous hydrodynamics model with finite baryon chemical potential combined with recent data from the RHIC beam energy scan.
By considering data from multiple beam energies, we can probe the temperature dependence of $\eta/s$.

We will include additional physical properties such as the size and shape of nucleons in the initial state, the hydrodynamic equation of state, and the switching temperature from hydrodynamics to microscopic transport; and compare to more observables, e.g.\ identified particle spectra and differential flow.

Finally, we anticipate upgrades to the methodology itself, notably more rigorous treatment of uncertainties and quantification of input-output correlations (analysis of variance).

\begin{acknowledgments}
We would like to thank Scott Pratt, Ulrich Heinz, Chun Shen, and Paul Sorensen for helpful discussions and valuable feedback.
This work would not have been possible without the foundations laid by the MADAI collaboration, funded through NSF grant no.~PHY-0941373, and the Statistical and Applied Mathematical Sciences Institute (SAMSI) program on massive datasets, funded through NSF grant no.~DMS-1127914.
SAB is supported by the U.S.\ Department of Energy grant no.~DE-FG02-05ER41367;
JEB by NSF grant no.~PHY-0941373 and DOE grant no.~DE-FG02-05ER41367;
RLW by NSF grant no.~DMS-1228317 and NASA AISR grant no.~NNX09AK60G.
This research was completed using over five million CPU hours provided by the Open Science Grid \cite{Pordes:2007zzb,Sfiligoi:2010zz}, which is supported by the National Science Foundation and the U.S.\ Department of Energy's Office of Science.
All code used in this project is publicly available \cite{Bernhard:2015mtd}.
Several third-party software packages were invaluable:
the Python MCMC toolkit \texttt{emcee} \cite{FM:2013mc} and Gaussian process library \texttt{george} \cite{Ambikasaran:2014gp}, and the general-purpose tool \texttt{GNU Parallel} \cite{Tange:2011pa}.
\end{acknowledgments}

\appendix

% only one appendix section -- don't want section a letter (Appendix A)
% manually set the equation numbering to (A1), (A2), ...
\section*{\label{app:train}Appendix: Training the emulator}
\renewcommand{\theequation}{A\arabic{equation}}
\setcounter{equation}{0}

\newcommand{\vectheta}{\boldsymbol\theta}

Gaussian process emulators are non-parametric models (they do not assume a functional form) but they do require an assumed covariance function.
One typically chooses a parameterized functional form based on physical considerations, for example the squared-exponential function
\begin{equation*}
  \sigma(\x, \x') = \exp\biggl( -\frac{|\x - \x'|^2}{2\ell^2} \biggr)
\end{equation*}
generates smoothly-varying processes and is therefore compatible with many models.
If the model is known to oscillate, one would choose a periodic covariance function.
Most models have statistical noise which one accounts for by adding a diagonal noise term to the covariance function:
\begin{equation*}
  \sigma(\x, \x') \rightarrow \sigma(\x, \x') + \sigma_n^2\delta_{\x\x'},
\end{equation*}
where $\sigma_n^2$ is the variance of the noise and $\delta_{\x\x'}$ is a Kronecker delta.

Covariance functions often have free parameters known as \emph{hyperparameters}, e.g.\ the squared-exponential correlation length $\ell$,
which are not known a priori and must be estimated from model data.
The selection of hyperparameters is known as \emph{training} and may be accomplished by maximizing the likelihood function \cite{Rasmussen:2006gp}
\begin{equation}
  \log P(\y|X,\vectheta) = -\frac{1}{2} \y\tran \Sigma_y^{-1} \y - \frac{1}{2} \log |\Sigma_y| - \frac{m}{2} \log 2\pi,
  \label{eq:train_like}
\end{equation}
where $\y$ is the vector of training outputs, $X$ is the matrix of training input points, $\vectheta$ is the vector of hyperparameters, and $\Sigma_y$ is the covariance matrix from applying the covariance function to the training data.
Hence, the likelihood is the probability of observing the data given the model.
The first term in the likelihood is the fit to data, the second term is a complexity penalty, and the third term is a normalization constant.

\colfig{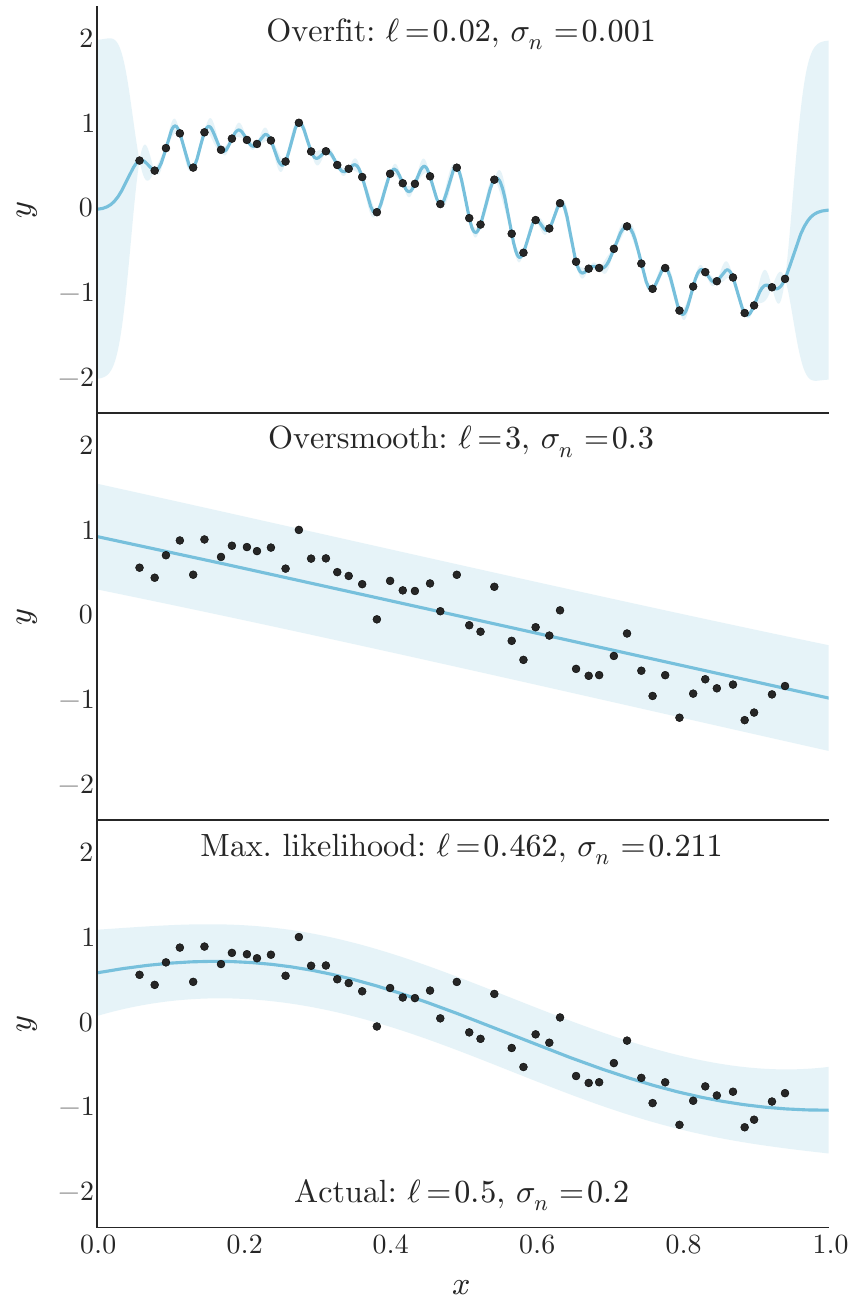}{
  Effect of varying the hyperparameters.
  Each panel shows a Gaussian process conditioned on fabricated training data using the covariance function Eq.~\eqref{eq:1dcov}, where the line is the mean and the band is a $2\sigma$ confidence interval.
  The covariance function hyperparameters are different in each plot as indicated by the annotations.
  The data points are identical in each plot and were generated from a Gaussian process with hyperparameters annotated at the bottom.
}

This is best demonstrated by an example.
The one-dimensional training data in Fig.~\ref{fig:training} were generated from a Gaussian process with covariance function
\begin{equation}
  \sigma(x, x') = \exp\biggl( -\frac{|x - x'|^2}{2\ell^2} \biggr) + \sigma_n^2\delta_{xx'}
  \label{eq:1dcov}
\end{equation}
and hyperparameters $\vectheta = (\ell, \sigma_n) = (0.5, 0.2)$; now let us pretend we don't know the true values of $\vectheta$ and attempt to train a Gaussian process on the data.
In the top panel of the figure, we use a short length scale $\ell$ with small noise $\sigma_n$, so the GP interpolates each point exactly; however it rapidly wiggles and is almost certainly ``overfit''.
This choice of hyperparameters has a large complexity penalty and therefore a low likelihood.
In the other extreme, we use a long length scale with large noise (middle panel), leading to a nearly linear GP that attributes most of the variance to noise.
Here the likelihood is also low due to the poor fit to data.
The most likely scenario is the compromise in the bottom panel, in which we estimate the hyperparameters by numerically maximizing the likelihood.
Now, the trained GP smoothly interpolates the curvature of the training data while leaving some of the variance as noise, true to the actual GP.

For the present study we use the covariance function given in Eq.~\eqref{eq:full_cov} and restated here:
\begin{equation*}
  \sigma(\x, \x') = \sigma_\text{GP}^2 \exp\Biggl[ -\sum_{k=1}^n \frac{(x_k - x'_k)^2}{2\ell_k^2} \Biggr] + \sigma_n^2\delta_{\x\x'}.
\end{equation*}
The hyperparameter $\sigma^2_\text{GP}$ is the overall variance of the Gaussian process and the $\ell_k$ are the independent length scales for each design parameter.
We estimate the hyperparameters by numerically maximizing the likelihood \eqref{eq:train_like} using a nonlinear conjugate gradient algorithm.
Since the likelihood may have non-optimal local extrema, we repeat the optimization algorithm many times (minimum 100) from different random starting points.

Table~\ref{tab:hyperpars} lists the maximum-likelihood estimates of the hyperparameters for each principal component in standardized units---input parameters scaled to $[0, 1]$ and principal components scaled to unit variance.
We constrain the length scales to $[0.3, 10]$ for numerical robustness.

In this work we fix the hyperparameters to the maximum-likelihood estimates during calibration.
This neglects uncertainty in the hyperparameters themselves, although the present event-by-event model is well-behaved and the sample size is large, so varying the hyperparameters weakly affects the actual emulator predictions.
But ideally, one would consider \emph{all} predictions consistent with the data---not only the most likely---by MCMC-sampling the hyperparameter posteriors.
This significantly increases computational cost, since the GPs must be reconditioned for every set of hyperparameters, and conditioning requires computation of the inverse covariance matrix, an $\mathcal O(n^3)$ operation.
We forgo this refinement until a future study.

\begin{table}[h]
  \caption{
    \label{tab:hyperpars}
    Maximum-likelihood estimates of the covariance function hyperparameters.
  }
  \begin{ruledtabular}
\begin{tabular}{llccccc}
  & & \multicolumn{5}{c}{Principal component} \\
  \noalign{\smallskip}\cline{3-7}\noalign{\smallskip}
  & & 1 & 2 & 3 & 4 & 5 \\
  \paddedhline
  \multirow{7}{*}{\rotatebox{90}{Glauber\ \ \ }}
  & $\sigma_\mathrm{GP}$ & 4.29 & 2.20 & 2.67 & 0.923 & 0.558 \\
  & $\ell$ Norm & 1.89 & 1.62 & 1.95 & 0.622 & 0.300 \\
  & $\ell$ $\alpha$ & 3.26 & 1.61 & 1.49 & 0.579 & 10.0 \\
  & $\ell$ $\tau_0$ & 1.20 & 1.05 & 1.65 & 1.34 & 0.300 \\
  & $\ell$ $\eta/s$ & 1.69 & 1.01 & 1.17 & 2.09 & 10.0 \\
  & $\ell$ $k_\pi$ & 10.0 & 4.77 & 4.46 & 10.0 & 1.48 \\
  & $\sigma_n$ & 0.0349 & 0.106 & 0.558 & 0.800 & 0.933 \\
  \paddedhline
  \multirow{7}{*}{\rotatebox{90}{KLN\ \ \ }}
  & $\sigma_\mathrm{GP}$ & 5.11 & 3.36 & 1.48 & 1.28 & 0.996 \\
  & $\ell$ Norm & 1.82 & 1.39 & 1.16 & 0.907 & 0.713 \\
  & $\ell$ $\lambda$ & 8.47 & 4.54 & 0.985 & 1.10 & 0.300 \\
  & $\ell$ $\tau_0$ & 0.927 & 0.678 & 0.808 & 0.534 & 0.359 \\
  & $\ell$ $\eta/s$ & 1.63 & 0.851 & 0.500 & 0.434 & 0.369 \\
  & $\ell$ $k_\pi$ & 10.0 & 8.33 & 2.04 & 1.43 & 0.389 \\
  & $\sigma_n$ & 0.0192 & 0.0568 & 0.807 & 0.803 & 0.606 \\
\end{tabular}
\end{ruledtabular}

\end{table}

\bibliography{mtd}

\end{document}